\documentclass[sigconf,nonacm]{sagepair-arxiv}

\usepackage{amsmath}
\usepackage{booktabs}
\usepackage{multirow}
\usepackage{xspace}

\newcommand{\PAIR}{\textsc{PAIR}\xspace}
\newcommand{\SAGE}{\textsc{SAGE-PAIR}\xspace}
\newcommand{\CoreSAGE}{\textsc{SAGE-PAIR} (Core)\xspace}
\newcommand{\FullSAGE}{\textsc{SAGE-PAIR} (Full)\xspace}

\newcommand{\R}{\mathbb{R}}
\newcommand{\softmax}{\operatorname{softmax}}
\newcommand{\BCE}{\operatorname{BCE}}

\renewcommand\footnotetextcopyrightpermission[1]{}
\begin{document}

\title{SAGE-PAIR: Source-Pretrained Routed Residual Adaptation for Sparse-Feedback Recommendation}

\author{Weiqing Zeng}
\affiliation{%
  \institution{Southwest Jiaotong University}
  \country{China}
}
\email{zengweiqing@my.swjtu.edu.cn}

\author{Zida Yang}
\affiliation{%
  \institution{Peking University}
  \country{China}
}
\email{2401111090@stu.pku.edu.cn}

\author{Guquan You}
\affiliation{%
  \institution{Guangzhou University}
  \country{China}
}
\email{youguquan@gzhu.edu.cn}

\author{Jiexin Zheng}
\affiliation{%
  \institution{Peking University}
  \country{China}
}
\email{jiexinzheng@gsm.pku.edu.cn}

\author{Cong Wang}
\authornote{Corresponding author.}
\affiliation{%
  \institution{Peking University}
  \country{China}
}
\email{wangcong@gsm.pku.edu.cn}
\renewcommand{\shortauthors}{Zeng et al.}

\begin{abstract}
Recommendation systems often enter a new serving window before enough outcomes are available to retrain the deployed model. We study this regime with $0.05$--$1.0\%$ target feedback, a range that spans our stress tests while keeping supervision far below the degrees of freedom of the source recommender. At this scale, existing fine-tuning and meta-learning methods still estimate target-specific updates from extremely sparse feedback, leaving the update direction weakly identified. \SAGE instead transfers a source-supported residual correction space and only calibrates this space using target outcomes. We introduce \SAGE, a staged adapter that pretrains a low-dimensional global residual on an adjacent source period and adds a soft router over residual experts conditioned on source user profiles. Target feedback then calibrates this residual state rather than relearning the recommender. On a locked test split of KuaiSAR-v2, \FullSAGE achieves a $1.77\%$ relative improvement in NDCG@10, while a core-protocol comparison shows a further $0.25\%$ gain from routed experts beyond source-pretrained \PAIR. Matched source-pretrained LoRA and Adapter-Tuning baselines remain near the source model, supporting the need for a different adaptation space. Thus, \SAGE improves ranking quality with millisecond-scale adaptation and inference costs. Sparse-feedback adaptation is fundamentally governed by the identifiability of the update space rather than by label count alone. Source pretraining supplies estimable correction directions for scarce labels, while user-conditioned routing contributes a consistent, differentiated cross-user correction on top of the global correction. Our code is available at \url{https://anonymous.4open.science/r/SAGE-PAIR-610E/}.
\end{abstract}



\maketitle

\section{Introduction}

Recommendation systems are frequently updated on a rolling schedule, but the evidence needed for a reliable update arrives more slowly than the serving schedule~\cite{koren2009temporal,wang2019temporal}. In particular, the shift from an adjacent source window (e.g., May 27) to a target window (e.g., May 28) can change user activity and item responses before enough outcomes are observed for retraining. This temporal drift can arise from daily changes in user activity, freshness- or event-driven item popularity, and delayed feedback that makes observed outcomes lag behind exposures~\cite{wang2019temporal,yang2021delayed}. We study this time-distribution-shift setting with an available source recommender, a fixed target exposure pool, and only a tiny fraction of target outcomes acquired before the next ranking update. The goal is to improve future user-grouped top-$k$ ranking while leaving the deployed backbone unchanged.

Achieving this goal requires estimating a deployable correction from far fewer target outcomes than the degrees of freedom of a recommender. Conventional adaptation procedures are mismatched to this constraint because they use the scarce labels to estimate updates in high-dimensional, target-specific parameter spaces. Full target fine-tuning updates user, item, and interaction parameters from the small target label set, so its update is weakly identified and can overwrite useful source structure. Parameter constraints such as L2-SP reduce this drift, but still optimize in the full backbone parameterization~\cite{li2018l2sp}. Adapter- and LoRA-style parameter-efficient tuning reduce the number of updated parameters by learning inserted or low-rank transformations~\cite{houlsby2019parameter,hu2022lora}; under our budget, however, they still estimate target-specific transformations across representation layers and leave the update directions weakly identified. MAML-style adaptation likewise relies on noisy task gradients in that high-dimensional space~\cite{finn2017model}. Cross-domain and cold-start methods transfer user representations or learn fast adaptation initializations~\cite{bharadhwaj2019coldstart,wang2021metatransitional,zhu2022ptupcdr,li2023oneforall,wu2023m2eu}; their objectives are complementary, but they do not directly address a fixed temporal exposure pool in which the deployable change is a very small residual around a frozen ranker.

The practical question is therefore not whether target feedback exists, but which correction directions it can support. A global residual is a robust answer because a few labels can estimate a small number of item-side directions. Yet a residual shared by every user cannot express that the same source-item direction may need different corrections for different source behavior regimes. Learning such user-specific corrections from target labels alone is precisely where scarce supervision is least reliable. The immediately preceding source window provides a distinct intermediate signal: unlike conventional domain adaptation, it uses labeled interactions from the same recommendation system and item schema to pretrain the correction space, while the target period only calibrates those source-supported directions.

Following this principle, we develop \SAGE as a modular framework for adaptation under temporal distribution shift. Its first module is a frozen-backbone residual adapter: a compact global correction uses source behavior statistics to define an update space while the deployed recommender remains unchanged. Its second module is staged source-to-target calibration: the residual is pretrained on the immediately preceding source window, then warm-started and calibrated with the sparse target outcomes, so the target labels estimate a small set of source-supported directions rather than a new recommender. Its third module is user-conditioned expert routing: a soft router maps stable source profiles and activity features to a mixture of residual experts, allowing the same global direction to receive differentiated corrections across user groups. These modules are optimized through the staged protocol and can be carried across successive target windows, providing a compact update path for continual serving.

The experiments separate the contributions of source pretraining and routed experts. On the locked KuaiSAR-v2 test split, \FullSAGE delivers a $1.77\%$ relative improvement in NDCG@10. In the core protocol, routed experts add a further $0.25\%$ relative gain beyond source-pretrained \PAIR, and continual warm-starting remains positive across repeated target windows.

Our contributions are:
\begin{itemize}
    \item \textbf{A staged frozen-backbone adaptation protocol.} We pretrain a compact global residual on an immediately preceding source window and calibrate only this residual with scarce target feedback, preserving the deployed recommender.
    \item \textbf{Source-pretrained routed residual experts.} We introduce a soft router and low-dimensional expert basis that condition residual corrections on source user profiles, enabling differentiated cross-user corrections within the same identifiable update space.
    \item \textbf{Continual warm-starting.} We carry the residual state across successive target windows while keeping the source profile as an anchor, improving repeated updates without retraining the backbone.
\end{itemize}

\section{Related Work}

\subsection{Sparse-Feedback Adaptation of Recommenders}

Matrix factorization, neural two-tower recommenders, and graph collaborative filtering provide different source representations for ranking~\cite{koren2009mf,covington2016youtube,he2020lightgcn}. When the target window is small, a direct update of these representations can have high variance; transfer constraints such as L2-SP instead preserve proximity to a source solution~\cite{li2018l2sp}. Adapter- and LoRA-style parameter-efficient tuning reduces update size by adding compact or low-rank trainable components~\cite{houlsby2019parameter,hu2022lora}, but still learns target-specific transformations from target feedback. \SAGE instead identifies a compact residual correction on the labeled adjacent source window, then uses the scarce target labels only to calibrate that correction. The backbone is frozen and only a two-dimensional source-behavior residual, its scale, and (for \SAGE) a small routed expert module are trained. This distinction matters because the same residual can be evaluated on future target rows without re-estimating user or item embeddings.

\subsection{Cross-Domain, Cold-Start, and Conditional Transfer}

Cross-domain recommendation transfers user preferences or user embeddings between domains~\cite{zhu2022ptupcdr,li2023oneforall}, while cold-start meta-learning learns an initialization that can be adapted from a few interactions~\cite{bharadhwaj2019coldstart,wang2021metatransitional,wu2023m2eu}. These approaches establish that source behavior can help sparse target personalization. \SAGE addresses a different transfer object and time structure: it transfers a residual ranking direction from an adjacent source period into a fixed target exposure protocol, rather than transferring a complete user representation or constructing a new meta-task for each user. The user profile conditions a shared expert basis, but the target update remains a small residual update around the same frozen source score.

Conditional computation and mixture-of-experts models use a gating function to combine specialized predictors~\cite{jacobs1991adaptive,ma2018mmoe}. \SAGE adopts this mechanism at the residual level, where its parameterization is small enough to be identifiable under the target budget. The relevant comparison is consequently not whether a router exists, but whether learned soft routing provides value beyond an equally source-pretrained global residual. Our core-vs-fallback comparison is designed to answer that question directly.

\subsection{Multi-Behavior Signals and Continual Updating}

Multi-behavior recommendation commonly combines auxiliary actions with a target behavior to model intent and preference~\cite{yan2024behavior}. \SAGE uses source click and like statistics as item-side residual features, providing a stable residual basis before the target window arrives. Continual and sequential cold-start methods similarly seek to update recommendations as new interactions appear~\cite{wang2021metatransitional}; \SAGE warm-starts the residual state across target windows while keeping the source profile fixed.

\subsection{Feedback Acquisition and Logged-Interaction Bias}

Active collaborative filtering and active matrix completion study which ratings or matrix entries to acquire under a limited budget~\cite{jin2004bayesian,harpale2008personalized,rubens2007influence,elahi2016survey,chakraborty2013active}. Batch active learning adds uncertainty and diversity objectives such as CoreSet and BADGE~\cite{settles2009active,sener2018coreset,ash2020badge}; recommender-specific work also optimizes item selection for exploration~\cite{kadiouglu2021optimized,kadiouglu2024integrating}. Counterfactual learning-to-rank and augmentation logging address exposure bias and evaluation variance~\cite{joachims2017unbiased,jagerman2019model,oosterhuis2021unifying,tucker2023mval}. Our primary \SAGE experiments use uniform outcome-blind selection, and we additionally evaluate CoreSet in the low-dimensional adapter feature space as an acquisition analysis.

\section{Problem Formulation}

\subsection{Temporal Adaptation Setting}

Each labeled interaction is a triple $(u,i,y)$ comprising user $u$, exposed item $i$, and binary click outcome $y\in\{0,1\}$. Historical labeled interactions $\mathcal{D}_0$ train the source recommender $f_{\theta}$, where $\theta$ denotes the backbone parameters. Labeled interactions from the immediately subsequent source window form $\mathcal{D}_1$; this window shares the same user, item, and behavior schema and is used to pretrain the adapter. The frozen source score for user--item pair $(u,i)$ is $z_{ui}=f_{\theta}(u,i)$.

The adjacent source window and the target window follow distributions $P_{\mathrm{src}}(u,i,y)$ and $P_{\mathrm{tgt}}(u,i,y)$, respectively. We study temporal distribution shift, defined as
\begin{equation}
P_{\mathrm{src}}(u,i,y)\neq P_{\mathrm{tgt}}(u,i,y).
\label{eq:temporal-shift}
\end{equation}
The shift may change exposed user--item pairs, their click propensity, or both.

\subsection{Sparse Target Feedback}

At the start of the target window, the system observes an outcome-blind exposure pool $\mathcal{Q}=\{q_r=(u_r,i_r)\}_{r=1}^{N}$ of $N$ target user--item pairs, but not their outcomes. An acquisition rule selects an index set $\mathcal{J}\subseteq\{1,\ldots,N\}$ of size $b=|\mathcal{J}|$, where $b$ is the feedback budget. The labeled adaptation set is $\mathcal{B}=\{(u_r,i_r,y_r):r\in\mathcal{J}\}$, so $|\mathcal{B}|=b$. A future held-out set $\mathcal{E}$ contains target-window interactions whose outcomes are never used in acquisition or fitting. Our stress tests use $b/N\in[0.0005,0.01]$; the locked KuaiSAR-v2 confirmation uses $b/N=0.005$.

\subsection{Objective and Staged Optimization}

Let $\phi\in\Phi$ denote the compact adapter parameters, where $\Phi$ is the low-dimensional adapter parameter space. The adapter defines an adapted score $s_{\phi}(u,i)=z_{ui}+\Delta_{\phi}(u,i)$, where $\Delta_{\phi}$ is a compact residual correction to the frozen source score. For a future user $u$, let $\mathcal{I}_u$ denote the candidate items in $\mathcal{E}$, $\mathbf{y}_u$ their binary outcomes, and $\mathbf{s}_{\phi}(u)=\{s_{\phi}(u,i):i\in\mathcal{I}_u\}$ their adapted scores. Let $\tau_u=(u,\mathcal{I}_u,\mathbf{y}_u)$ denote this user-level evaluation instance, and let $P_{\mathrm{tgt}}^{\mathrm{user}}$ be the user-level distribution induced by $P_{\mathrm{tgt}}$. With ranking cutoff $k$, the deployment objective is
\begin{equation}
\max_{\phi\in\Phi}\ \mathbb{E}_{\tau_u\sim P_{\mathrm{tgt}}^{\mathrm{user}}}\!\left[\mathrm{NDCG@}k\bigl(\mathbf{y}_u,\mathbf{s}_{\phi}(u)\bigr)\right].
\label{eq:problem-objective}
\end{equation}

The held-out set $\mathcal{E}$ estimates this target expectation. The feasible adapter is obtained by pretraining on $\mathcal{D}_1$ and warm-updating on $\mathcal{B}$. The backbone parameters $\theta$, source-derived item features $\{\mathbf{x}_i\}$, and source user profiles $\{\mathbf{h}_u\}$ remain fixed; only $\phi$ is updated, using the $b$ revealed target labels.

Because the outcomes in $\mathcal{E}$ are unavailable during adaptation, optimization uses a regularized binary cross-entropy surrogate. For any labeled set $\mathcal{T}$, where each element is an interaction triple, the loss is
\begin{equation}
\mathcal{L}(\phi;\mathcal{T})=\frac{1}{|\mathcal{T}|}\sum_{(u,i,y)\in\mathcal{T}}\BCE\bigl(s_{\phi}(u,i),y\bigr)+\Omega(\phi).
\label{eq:adapter-surrogate}
\end{equation}
Here, $\BCE$ is binary cross-entropy and $\Omega(\phi)$ is an L2-type penalty on the residual parameters that anchors the update near the source score. Source pretraining minimizes $\mathcal{L}(\phi;\mathcal{D}_1)$ to obtain $\phi_{\mathrm{src}}$:
\begin{equation}
\phi_{\mathrm{src}}=\arg\min_{\phi\in\Phi}\mathcal{L}(\phi;\mathcal{D}_1).
\label{eq:source-pretrain-objective}
\end{equation}
Starting from $\phi_{\mathrm{src}}$, the target warm update minimizes $\mathcal{L}(\phi;\mathcal{B})$ and returns the final adapter $\widehat{\phi}$. Section~\ref{sec:method-global} specifies the residual score and regularizer, and Section~\ref{sec:method-sage} extends them with routed experts.

\section{Method}

\subsection{Overview}

\begin{figure*}[t]
\centering
\includegraphics[width=\textwidth]{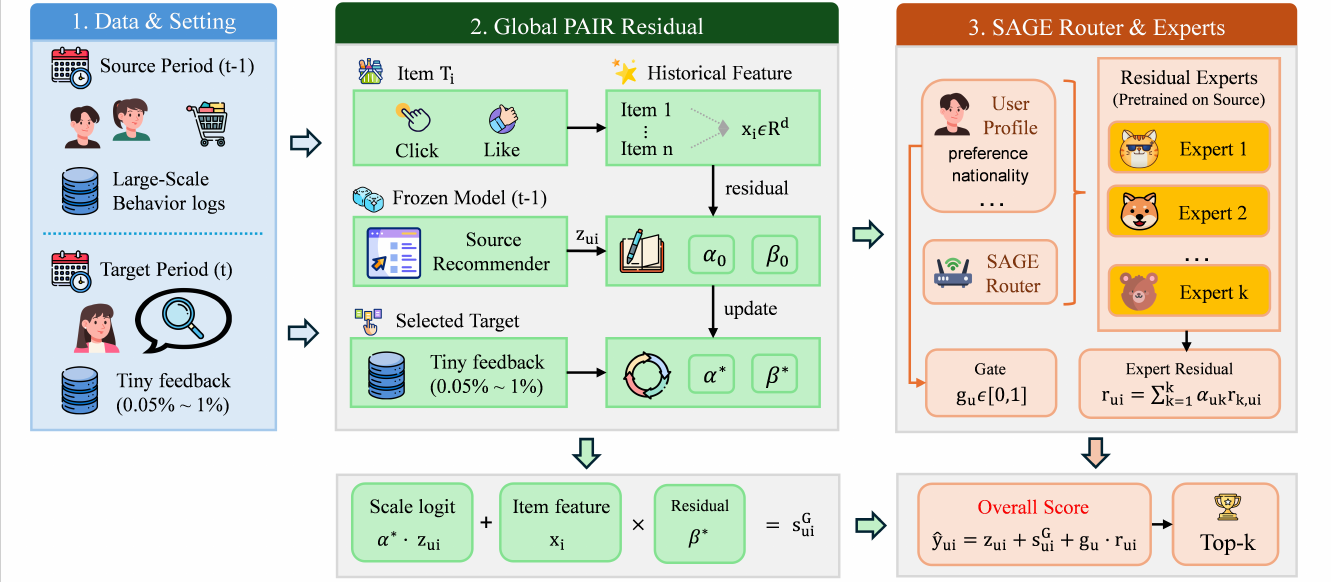}
\caption{Architecture of \SAGE. The source recommender remains frozen and supplies the base score, while source-derived item behavior features and user profiles parameterize a compact global residual and user-routed residual experts. These residual components are pretrained on an adjacent source window and warm-updated using the outcome-blind target feedback budget before scoring future target rows.}
\Description{The \SAGE architecture has a frozen source recommender and a compact residual adaptation module. Source item behavior features and user profiles support a global residual and user-routed experts, which are pretrained on an adjacent source window and warm-updated with selected target feedback before scoring future target rows.}
\label{fig:pipeline}
\end{figure*}

\SAGE is a staged residual adapter around a frozen source recommender. Figure~\ref{fig:pipeline} summarizes the architecture and training stages. Source-period interactions train the backbone and yield fixed item behavior features and user profiles. An adjacent source window pretrains the global residual and the routed residual experts. After outcome-blind target query selection, only this compact residual module is warm-updated; the frozen backbone then combines its original score with the adapted residual to rank future target rows.

\subsection{Source-Only Item Behavior Features}

The residual uses source behavior statistics rather than a new target-trained item representation. For item $i$, behavior $k$, exposure count $n_i$, positive count $c_{ik}$, and global source rate $\bar p_k$, we compute an empirical-Bayes rate
\[
\widehat{p}_{ik} =
\frac{c_{ik}+\alpha\bar p_k}{n_i+\alpha},
\qquad
x_{ik} =
\frac{\widehat{p}_{ik}-\mu_k}{\sigma_k},
\]
where $\alpha=20$ in the selected click--like setting and $(\mu_k,\sigma_k)$ are source-catalog moments. The resulting $\mathbf{x}_i\in\R^d$ is fixed before target labels are accessed. In the main protocol, $d=2$ because the two coordinates are exactly the source click-rate and like-rate statistics; it is not a learned projection of a higher-dimensional target residual. This feature choice deliberately restricts adaptation to reweighting two source-estimable behavior signals, rather than asking sparse target labels to identify an unconstrained item representation. Shrinkage prevents rare like events from producing an uncontrolled residual direction, while standardization makes the regularization comparable across behaviors.

\subsection{Global PAIR and Source Residual Pretraining}
\label{sec:method-global}

The global \PAIR adapter is
\begin{equation}
s^{\mathrm{G}}_{ui}
=a z_{ui}+\mathbf{x}_i^\top\boldsymbol{\beta},
\qquad
\boldsymbol{\omega}_{\mathrm{G}}=(a,\boldsymbol{\beta}).
\label{eq:global}
\end{equation}
The scalar $a$ permits a small global recalibration of the frozen source score under temporal shift, while $\boldsymbol{\beta}$ weights the source-only behavior features. The penalty $\lambda_a(a-1)^2$ anchors $a$ near the source calibration: it permits a confidence adjustment when needed, but prevents the adapter from using a free score rescaling as a substitute for retraining the backbone.
Target-only \PAIR estimates $\boldsymbol{\omega}_{\mathrm{G}}$ from the selected target rows $\mathcal{B}$. Its staged version, source-pretrained \PAIR, first estimates the same parameters on $\mathcal{D}_1$, then continues optimization on $\mathcal{B}$. It retains the same scoring form without a user-conditioned module.

Let $\boldsymbol{\omega}$ denote the active adapter state: $\boldsymbol{\omega}_{\mathrm{G}}=(a,\boldsymbol{\beta})$ for global \PAIR, and $\boldsymbol{\omega}_{\mathrm{S}}=(a,\boldsymbol{\beta},W,R)$ for \CoreSAGE (defined below). Let $s_{\boldsymbol{\omega}}(u,i)$ denote the corresponding score. The two-stage objective is
\begin{equation}
\mathcal{L}(\boldsymbol{\omega};\mathcal{T})
=\frac{1}{|\mathcal{T}|}\sum_{(u,i,y)\in\mathcal{T}}
\BCE(s_{\boldsymbol{\omega}}(u,i),y)
+\lambda_\beta\|\boldsymbol{\beta}\|_2^2
+\lambda_a(a-1)^2,
\label{eq:global-loss}
\end{equation}
with $\lambda_\beta=0.02$ and $\lambda_a=0.01$. Source pretraining changes the information available to the residual, not the backbone: $f_\theta$ remains frozen in both stages. The target update therefore calibrates a source-derived direction instead of estimating a high-dimensional recommendation model from scratch.

\subsection{\CoreSAGE: Source-Pretrained Soft Routed Experts}
\label{sec:method-sage}

The global residual is reliable but user agnostic. \CoreSAGE has parameter space $\Phi_{\mathrm{Core}}=\{a,\boldsymbol{\beta},W,R\}$ and augments the global state $\boldsymbol{\omega}_{\mathrm{G}}$ with the routing parameters $W$ and $R$, so its complete parameter state is $\boldsymbol{\omega}_{\mathrm{S}}=(a,\boldsymbol{\beta},W,R)$. It adds a small, source-profile-conditioned residual basis. Let $\mathbf{h}_u\in\R^p$ denote a source user profile containing a normalized recommendation-category click histogram and log activity summaries. This is an explicit descriptor rather than a learned $p$-dimensional embedding. On KuaiSAR-v2, $p=39$ comprises 37 first-level-category histogram bins and two activity slots. The core protocol populates the histogram from recommendation behavior and retains the search-activity slot at zero; the fixed tensor shape makes the search-profile ablation a controlled input change rather than a change in router capacity.

The router uses a learned matrix $W\in\R^{K\times p}$ to map a profile to a probability simplex:
\begin{equation}
\boldsymbol{\pi}_u=\softmax(W\mathbf{h}_u),
\qquad
\boldsymbol{\pi}_u\in\R^K,\quad
\sum_{k=1}^{K}\pi_{uk}=1.
\label{eq:router}
\end{equation}
The $K$ rows of $R\in\R^{K\times d}$ are residual coefficient anchors. The core score is
\begin{equation}
s^{\mathrm{SAGE\text{-}PAIR}}_{ui}
=a z_{ui}
+\mathbf{x}_i^\top\boldsymbol{\beta}
+\mathbf{x}_i^\top(\boldsymbol{\pi}_u^\top R).
\label{eq:sage}
\end{equation}
The first two terms are the staged global PAIR component; the last term is the user-conditioned increment. Although the expert term is linear in $\mathbf{x}_i$ for a fixed user, the coefficient $\boldsymbol{\pi}_u^\top R$ is a nonlinear, profile-dependent function through the softmax router; it is therefore not a single shared linear residual. The experts are not intended to be mutually orthogonal full predictors: with $K=8$ and $d=2$, $\operatorname{rank}(R)\leq2$, so a penalty that forces $RR^\top$ toward the $K\times K$ identity $I_K$ is infeasible. Instead, the design uses an overcomplete set of low-dimensional coefficient anchors and tests whether learned routing contributes beyond the equally source-pretrained global residual. Soft routing lets a user share several residual regimes rather than committing to a hard cluster, while the low-dimensional item basis keeps each user-conditioned correction in the same estimable behavior space.

The locked \FullSAGE confirmation additionally multiplies the expert term by $g_u=\operatorname{sigmoid}(v^\top\mathbf{h}_u)$, where $v\in\R^p$ is a learned gate vector, and includes search-profile channels; its parameter space is therefore $\Phi_{\mathrm{Full}}=\{a,\boldsymbol{\beta},W,R,v\}$. The selected \CoreSAGE uses source pretraining, learned routing, and $K=8$ experts. The ablation results show that source pretraining and learned routing drive the main gain, while search and gating have small effects.

\subsection{Staged Optimization and Deployment Fallback}

For source pretraining, the backbone score and source-only features are computed first, and $(W,R,\boldsymbol{\beta},a)$ are optimized on $\mathcal{D}_1$ for five epochs with Adam, learning rate $0.02$, and weight decay $10^{-4}$. The expert loss adds
\[
0.1\|R\|_F^2
\]
to Equation~\eqref{eq:global-loss}. The staged transfer is an initialization, not a constraint: during target adaptation, every residual parameter is copied from the adjacent source state and then optimized for eight epochs with learning rate $0.01$, the same weight decay, and the selected target rows only. Thus, source pretraining supplies a data-informed starting direction while target feedback remains the sole fitting signal in the target stage. Using the immediately preceding window keeps the scoring task, catalog schema, and serving context fixed while providing labels that are unavailable at the start of the target window. The source backbone and all source feature construction remain unchanged.

The deployment fallback is explicit. When a source profile and pretrained router are available, the system uses \CoreSAGE. If the required source profile is unavailable, it drops the expert term and uses source-pretrained \PAIR. If no adjacent source residual window exists, target-only \PAIR remains available. This hierarchy makes the extra module additive: it cannot force a user onto a newly estimated high-dimensional model.

\subsection{Continual Warm-Start}
\label{sec:method-continual}

Let $\boldsymbol{\omega}_t$ denote the residual state after target window $t$. Given a new outcome-blind sample $\mathcal{B}_{t+1}$, continual \SAGE updates
\begin{equation}
\boldsymbol{\omega}_{t+1}
=\operatorname{Update}(\boldsymbol{\omega}_t,\mathcal{B}_{t+1}),
\label{eq:continual}
\end{equation}
where $\operatorname{Update}$ is the same residual optimizer used for a target warm update. The source profile $\mathbf{h}_u$ and source item features $\mathbf{x}_i$ remain fixed. A daily-reset baseline instead initializes every window from the source-pretrained state. We also evaluate a profile-refresh variant that adds sparse daily target behavior to $\mathbf{h}_u$; this variant is not part of the proposed method.

\subsection{Parameterization and Complexity}

For $p=39$, $d=2$, and $K=8$, the ablation-compatible implementation stores
\[
 Kp+(p+1)+Kd+d+1=371
\]
scalar parameter slots. The $p+1$ gate slots are retained for shape-compatible comparisons with the full gated variant but do not enter the core score; \CoreSAGE therefore uses $Kp+Kd+d+1=331$ active scoring parameters. Source expert pretraining takes $0.577$ seconds and the instrumented CUDA peak allocation is $83.15$ MB. The key design property is that adaptation cost is independent of the backbone parameter count.

\section{Experimental Setup}

\subsection{Datasets and Temporal Protocol}

We evaluate the method in rolling temporal protocols in which the source recommender is trained before the target period and is never refit during adaptation. The broad PAIR study uses KuaiRand, KuaiRec, and Tenrec, with an additional KuaiRec LightGCN run to test dependence on the source backbone. KuaiRand provides randomly exposed videos and is useful for reducing the confounding between exposure and outcome; KuaiRec provides a fully observed recommendation setting; and Tenrec supplies a separate large-scale multi-purpose recommendation corpus~\cite{gao2022kuairand,gao2022kuairec,yuan2022tenrec}. The same source-only feature construction and outcome-blind selection rule are used within each protocol. Table~\ref{tab:dataset-statistics} reports the processed protocol sizes rather than corpus-level headline counts.

The main staged-method evaluation is KuaiSAR-v2. The source backbone uses interactions from May 22--26, source residual pretraining uses May 27, the target query budget is selected on May 28, and evaluation uses future outcomes from May 29--31. The target query contains $0.5\%$ of the eligible target pool, and labels are hidden until the query rows have been fixed. KuaiSAR-v2 has only nine user IDs in common with the KuaiSAR-small development population, providing a near-disjoint-user confirmation of source transfer. The locked \FullSAGE confirmation uses five source seeds and five selection seeds, for 25 paired cells. The fair \CoreSAGE decomposition uses three source seeds and five selection seeds, for 15 cells; both comparisons use the same temporal split and target query budget within each cell.

\begin{table}[t]
  \caption{Processed protocol statistics. User and item counts are distinct IDs in backbone-source data. S/Q/E rows (M) denote source, target query pool, and evaluation interactions. KuaiSAR-v2 additionally uses 1.23M May~27 pretraining rows and acquires 5.8K May~28 outcomes.}
  \label{tab:dataset-statistics}
  \centering
  \small
  \setlength{\tabcolsep}{3pt}
  \renewcommand{\arraystretch}{1.06}
  \begin{tabular}{lrr}
    \toprule
    Dataset & Users / Items & S / Q / E (M) \\
    \midrule
    KuaiRand-Pure & 26.2K / 7.5K & 1.14 / 0.86 / 0.22 \\
    KuaiRec 2.0 & 7.2K / 10.7K & 12.53 / 2.34 / 2.34 \\
    Tenrec-QB-video & 34.2K / 104.4K & 1.47 / 0.47 / 0.46 \\
    KuaiSAR-v2 & 17.2K / 1.25M & 3.37 / 1.16 / 2.26 \\
    \bottomrule
  \end{tabular}
\end{table}

\subsubsection{Drift Quantification}
\label{sec:drift-quantification}

We quantify the adjacent May~27 $\rightarrow$ May~28 transition rather than inferring drift from calendar proximity. The item-popularity distribution has Jensen--Shannon (JS) divergence $0.381048$ across the two windows; only $29.0\%$ of May~28 items also occur on May~27. User turnover is also substantial: only $48.8\%$ of May~28 users overlap with May~27. In contrast, the first-level-category and log activity histograms are comparatively stable, which localizes the shift to item and user identities rather than a broad category-mix change.

We additionally evaluate our source Two-Tower model, trained on May~22--26 and not adapted on either May~27 or May~28. Across three source seeds, Macro-AUC falls from $0.553135$ to $0.546759$ and binary cross-entropy (BCE) rises from $0.666710$ to $0.692399$. NDCG@10 increases slightly ($0.644807$ to $0.647561$), which is expected to be sensitive to the changed daily candidate sets and click prevalence; we therefore do not characterize the shift as an NDCG drop. Together, the identity-level covariate shift and source-model discrimination/calibration degradation provide the empirical evidence for $P_{\mathrm{src}} \ne P_{\mathrm{tgt}}$.

\begin{table*}[t]
  \caption{May~27 $\rightarrow$ May~28 temporal-shift audit on KuaiSAR-v2. JS is Jensen--Shannon divergence. Our source Two-Tower model is trained on May~22--26 and evaluated without May~27 or May~28 adaptation; metrics are means over three source seeds.}
  \label{tab:drift-audit}
  \centering
  \small
  \renewcommand{\arraystretch}{1.06}
  \begin{tabular}{lrrl}
    \toprule
    Quantity & May 27 & May 28 & Shift statistic \\
    \midrule
    Interactions & 1,227,417 & 1,164,683 & -- \\
    Click rate & 0.518963 & 0.524632 & $+0.005669$ \\
    Unique users & 10,122 & 12,443 & 48.8\% May~28 coverage; Jaccard 0.3680 \\
    Unique items & 482,929 & 465,784 & 29.0\% May~28 coverage; Jaccard 0.1662 \\
    Item-popularity distribution & -- & -- & JS $=0.381048$ \\
    First-level-category distribution & -- & -- & JS $=0.000309$ \\
    Log user-activity distribution & -- & -- & JS $=0.004112$ \\
    Our source model: NDCG@10 & 0.644807 & 0.647561 & $+0.002754$ \\
    Our source model: Macro-AUC & 0.553135 & 0.546759 & $-0.006376$ \\
    Our source model: BCE & 0.666710 & 0.692399 & $+0.025689$ \\
    \bottomrule
  \end{tabular}
\end{table*}

\subsection{Backbones, Methods, and Baselines}

The default source backbone is a two-tower recommender. We additionally evaluate matrix factorization (MF) and LightGCN to separate the adapter effect from a particular backbone parameterization~\cite{he2020lightgcn}. All methods receive the same frozen source score and the same future evaluation rows. Target fine-tuning, L2-SP, and source replay update the full backbone and serve as budget-matched controls. The global \PAIR baseline updates only the low-dimensional residual from target feedback. Its staged counterpart, source-pretrained \PAIR, receives exactly the same adjacent source residual window as \SAGE but has no user-conditioned expert term.

To make the parameter-efficiency comparison external rather than solely ablative, we compare with target-only LoRA and bottleneck Adapter-Tuning on the source Two-Tower item projection~\cite{houlsby2019parameter,hu2022lora}. Both have rank four and exactly 256 trainable parameters, close to \CoreSAGE's 331 active scoring parameters. We further include first-order MAML (FOMAML)~\cite{finn2017model} over the same rank-four LoRA factors: its initialization is meta-trained on disjoint support/query episodes drawn within the five chronological source days and then updated only on the selected May~28 labels. Finally, we run source-pretrained LoRA and source-pretrained Adapter-Tuning, which use the May~27 adjacent source window for five epochs before the same May~28 warm update for eight epochs. All variants use no target outcomes beyond the shared outcome-blind query budget; ranks, learning rates, epochs, source seed blocks, selection seeds, and future rows are fixed to the common KuaiSAR-v2 protocol.

For the main method, we report both the locked full variant and the selected core variant. \FullSAGE includes search-profile channels and a scalar reliability gate in the end-to-end confirmation. \CoreSAGE excludes both additions while retaining source residual pretraining, a learned soft router, and $K=8$ low-dimensional residual experts. The core comparison pairs \CoreSAGE with source-pretrained \PAIR so that both methods receive the same adjacent source window. Cohort baselines form item-category- and user-preference-conditioned groups from source behavior. They are included as strong structured baselines, not as components of \SAGE.

\subsection{Metrics and Statistical Reporting}

We report user-grouped NDCG@10, Recall@10, NDCG@20, Recall@20, and macro-averaged per-user AUC. Every metric is grouped by the raw \texttt{user\_id}; an internal index used by the backbone for an unseen entity is never used as a metric group. The main comparison tables report absolute metric means with methods as rows and datasets or metrics as columns; the best method within each protocol block is shown in bold. Relative percentage increases are discussed in the text when useful. Differences are paired within each source-seed/selection-seed cell. Unless otherwise stated, confidence intervals are 95\% intervals over paired cells, and the positive-cell count reports how many cells have a positive difference. For the headline result, we also report the 95\% interval over source-seed means, because selection seeds are nested within each source seed. This reporting makes the distinction between repeated selection variability and independent source-model variability explicit.

\subsection{Implementation and Reproducibility}

The source profile has dimension $p=39$, the source item residual has dimension $d=2$, and the selected expert count is $K=8$. The frozen backbone uses AdamW with learning rate $0.004$ and weight decay $10^{-6}$. Source residual pretraining uses Adam with learning rate $0.02$, five epochs, and weight decay $10^{-4}$; target warm updates use learning rate $0.01$, eight epochs, and the same weight decay. These settings are fixed across the main KuaiSAR comparisons. \CoreSAGE has 331 active scoring parameters; the recorded ablation-compatible implementation stores 371 slots because it retains the inactive gate tensor. In the recorded server run, source expert pretraining took $0.577$ seconds and the instrumented CUDA peak allocation was $83.15$ MB.

The epoch counts are selected before the final May~28--31 confirmation by a strictly earlier rolling development protocol: the backbone trains on May~22--24, the source residual trains on May~25, a $0.5\%$ outcome-blind query is drawn from May~26, and May~27 supplies validation outcomes. We sweep source epochs $\{1,3,5,8\}$ and target epochs $\{1,3,5,8,12\}$ over three source seeds and three selection seeds, without reading May~28--31 labels. The maximum mean validation NDCG@10 is attained at five source and eight target epochs ($0.662009$); twelve target epochs plateau at $0.662002$ rather than improving the result. We then freeze these epoch counts for every final comparison. We use uniform outcome-blind selection for the primary experiments and evaluate active acquisition separately.

\section{Results}

\subsection{Broad Confirmation of the Global Residual}

We first establish the behavior of the low-dimensional residual independently of routed experts. Table~\ref{tab:pair-broad} summarizes the locked PAIR confirmations. PAIR raises NDCG@10 from $0.399169$ to $0.459226$ on KuaiRand (a $15.05\%$ relative increase), from $0.919298$ to $0.924952$ on KuaiRec ($0.62\%$), from $0.859586$ to $0.862950$ on Tenrec ($0.39\%$), and from $0.920140$ to $0.926536$ with LightGCN ($0.70\%$); every paired cell is positive. Thus, the residual parameterization is not tied to one source backbone or one dataset exposure process.

\begin{table}[t]
  \caption{Broad confirmation of the global residual. Entries are user-grouped NDCG@10 $\uparrow$ means; TT denotes the Two-Tower backbone.}
  \label{tab:pair-broad}
  \centering
  \small
  \renewcommand{\arraystretch}{1.06}
  \begin{tabular}{lrrr}
    \toprule
    Setting & \multicolumn{2}{c}{NDCG@10 $\uparrow$} & Rel. gain $\uparrow$ \\
    \cmidrule(lr){2-3}
    & Source & \PAIR & \\
    \midrule
    KuaiRand / MF+TT & 0.399169 & \textbf{0.459226} & 15.05\% \\
    KuaiRec / MF+TT & 0.919298 & \textbf{0.924952} & 0.62\% \\
    Tenrec / MF+TT & 0.859586 & \textbf{0.862950} & 0.39\% \\
    KuaiRec / LightGCN & 0.920140 & \textbf{0.926536} & 0.70\% \\
    \bottomrule
  \end{tabular}
\end{table}

The same pattern holds under very small budgets. On KuaiSAR-small, click-only PAIR raises NDCG@10 from $0.626024$ to $0.647852$ at $0.1\%$ feedback (a $3.49\%$ relative increase) and to $0.647340$ at $0.5\%$ ($3.40\%$), both positive in all 25 cells. The corresponding Two-Tower score rises from $0.629072$ to $0.640034$ at $0.1\%$ ($1.74\%$) and from $0.629072$ to $0.642520$ at $0.5\%$ ($2.14\%$). Full-model target fine-tuning remains essentially unchanged, from $0.626024$ to $0.626022$ and from $0.626024$ to $0.626048$ in the same sparse-feedback setting. These controls support the specific claim that the useful adaptation direction is low-dimensional; simply exposing the full backbone to the same labels does not recover it.

\subsection{Locked KuaiSAR-v2 Confirmation}

Table~\ref{tab:multimetric-confirmation} reports the frozen five-metric completion on the near-disjoint KuaiSAR-v2 population. In the locked 25-cell confirmation, \FullSAGE raises NDCG@10 from $0.615217$ to $0.626115$ (a $1.77\%$ relative increase), Recall@10 from $0.278570$ to $0.280224$ ($0.59\%$), NDCG@20 from $0.644060$ to $0.651995$ ($1.23\%$), Recall@20 from $0.424616$ to $0.426032$ ($0.33\%$), and Macro-AUC from $0.548855$ to $0.552616$ ($0.69\%$). The NDCG@10 interval is $[+0.010054,+0.011740]$, and all 25 paired cells are positive. The improvement is therefore not a single-cutoff artifact: all five metrics move in the same direction.

\begin{table*}[t]
  \caption{KuaiSAR-v2 results at a $0.5\%$ outcome-blind target-feedback budget. Parenthetical labels identify the locked and core protocols; bold denotes the better method within each comparison.}
  \label{tab:multimetric-confirmation}
  \centering
  \small
  \renewcommand{\arraystretch}{1.06}
  \begin{tabular}{lrrrrr}
    \toprule
    Method & NDCG@10 $\uparrow$ & Recall@10 $\uparrow$ & NDCG@20 $\uparrow$ & Recall@20 $\uparrow$ & Macro-AUC $\uparrow$ \\
    \midrule
    Target-only \PAIR (locked) & 0.615217 & 0.278570 & 0.644060 & 0.424616 & 0.548855 \\
    \textsc{SAGE-PAIR} (Full; locked) & \textbf{0.626115} & \textbf{0.280224} & \textbf{0.651995} & \textbf{0.426032} & \textbf{0.552616} \\
    \midrule
    Target-only \PAIR (core) & 0.624800 & 0.281265 & 0.652100 & 0.426721 & 0.553421 \\
    Source-pretrained \PAIR (core) & 0.631008 & 0.282253 & 0.656593 & 0.427605 & 0.555788 \\
    \CoreSAGE & \textbf{0.632608} & \textbf{0.282614} & \textbf{0.657802} & \textbf{0.427942} & \textbf{0.556573} \\
    \bottomrule
  \end{tabular}
\end{table*}

\subsection{External PEFT and Meta-Learning Comparison}

The external comparison supports the staged adaptation claim, with an important qualification. Table~\ref{tab:external-peft} evaluates target-only rank-four LoRA, rank-four bottleneck Adapter-Tuning, FOMAML-LoRA, and source-pretrained variants under the same $0.5\%$ outcome-blind target budget and the core evaluation protocol. Target-only Adapter-Tuning gives a small positive NDCG@10 gain of $+0.000462$; its 95\% interval is above zero, while the LoRA and FOMAML intervals include zero.

The source-pretrained LoRA and Adapter-Tuning variants decrease NDCG@10 by $0.000372$ and $0.000553$, respectively. Their full intervals and positive-cell counts are reported in Table~\ref{tab:external-peft}. Thus, these results do \emph{not} justify claiming that all parameter-efficient methods are ineffective; they show that direct target-only PEFT is modest, whereas naively pretraining the same low-rank or bottleneck modules on the adjacent source window is not reliable under this shift.

These results explain why source pretraining alone is insufficient when applied as a generic representation-space transformation. Source-pretrained LoRA and Adapter-Tuning learn transformations in the backbone representation space, whereas \SAGE pretrains the correction directly in the observable, source-supported residual space. Under the extremely sparse target budget, the former still estimates transformation parameters whose target relevance is weakly identified; the latter restricts adaptation to empirically supported item-side directions.

By contrast, source-pretrained \PAIR and \CoreSAGE improve over the source model by $+0.028814$ and $+0.030407$, respectively. \CoreSAGE exceeds source-pretrained LoRA and Adapter-Tuning by $+0.030779$ and $+0.030960$, with all 15 paired cells positive. It also exceeds the strongest target-only baseline, Adapter-Tuning, by $+0.029944$ NDCG@10. These comparisons are external to the \PAIR ablation; the source-pretrained \PAIR row separately measures the additional contribution of routed experts.

\begin{table*}[t]
  \caption{External parameter-efficient and meta-learning baselines on KuaiSAR-v2. The key staged ordering is source-pretrained LoRA/Adapter-Tuning $\ll$ source-pretrained \PAIR $<$ \CoreSAGE. All methods use the same $0.5\%$ outcome-blind May~28 budget and May~29--31 evaluation; CIs are paired 95\% intervals over 15 source-seed/selection-seed cells. LoRA, Adapter-Tuning, and FOMAML each tune 256 parameters; \CoreSAGE uses 331 active scoring parameters.}
  \label{tab:external-peft}
  \centering
  \small
  \renewcommand{\arraystretch}{1.06}
  \begin{tabular}{lrrrr}
    \toprule
    Method & NDCG@10 $\uparrow$ & $\Delta$ vs. source & 95\% CI of $\Delta$ & Positive cells \\
    \midrule
    Our source model & 0.602014 & -- & -- & -- \\
    Target-only LoRA (rank 4) & 0.602186 & +0.000172 & [-0.000058, +0.000401] & 11/15 \\
    Target-only Adapter-Tuning (rank 4) & 0.602476 & +0.000462 & [+0.000072, +0.000852] & 12/15 \\
    FOMAML-LoRA (rank 4) & 0.602204 & +0.000189 & [-0.000055, +0.000434] & 12/15 \\
    Source-pretrained LoRA (rank 4) & 0.601642 & -0.000372 & [-0.000726, -0.000019] & 3/15 \\
    Source-pretrained Adapter-Tuning (rank 4) & 0.601461 & -0.000553 & [-0.000987, -0.000120] & 4/15 \\
    Source-pretrained \PAIR & 0.630828 & +0.028814 & [+0.027035, +0.030593] & 15/15 \\
    \CoreSAGE & \textbf{0.632421} & \textbf{+0.030407} & \textbf{[+0.028807, +0.032006]} & \textbf{15/15} \\
    \bottomrule
  \end{tabular}
\end{table*}

The unadapted source model scores $0.593792$ NDCG@10, while target-only \PAIR and \FullSAGE reach $0.615217$ and $0.626115$, respectively. The decomposition is the central result of the paper.

In the fair 15-cell core protocol, source-pretrained \PAIR raises NDCG@10 from $0.624800$ to $0.631008$ (a $0.99\%$ relative increase). \CoreSAGE raises it further to $0.632608$ (an additional $0.25\%$). The same ordering appears for Recall@10, NDCG@20, Recall@20, and Macro-AUC. Source pretraining is consequently the main stabilizer; routed experts add a smaller but consistent source-profile-conditioned correction.

\begin{figure*}[t]
  \centering
  \includegraphics[width=\textwidth]{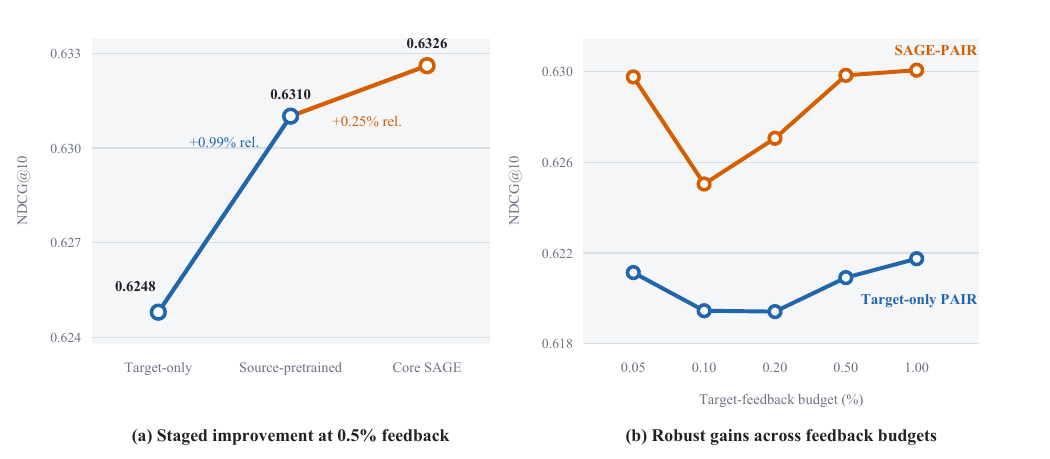}
  \caption{Staged gain decomposition and feedback-budget robustness. (a) In the $0.5\%$ core protocol, source pretraining yields a $0.99\%$ relative NDCG@10 gain over target-only \PAIR, while routed experts add a further $0.25\%$ relative gain. (b) \SAGE exceeds target-only \PAIR at every target-feedback budget from $0.05\%$ to $1.0\%$.}
  \Description{Two line charts. The left panel decomposes the NDCG@10 improvement at 0.5 percent feedback into the gain from source pretraining and the additional gain from user-conditioned routing. The right panel plots NDCG@10 over five feedback budgets and shows SAGE-PAIR above target-only PAIR at every budget.}
  \label{fig:result-metrics}
\end{figure*}

Figure~\ref{fig:result-metrics} visualizes this decomposition together with the feedback-budget curve. The two views distinguish the large source-pretraining gain from the additional routing gain and show that the advantage persists throughout the sparse-feedback regime.

\subsection{Model Selection}

Table~\ref{tab:sage-ablation-main} isolates the design choices on the KuaiSAR-v2 core protocol. Against the $0.619675$ target-only \PAIR baseline, \FullSAGE reaches $0.627783$ and \CoreSAGE reaches $0.628131$, a $1.36\%$ relative increase. Removing the search profile gives $0.627694$, while removing the scalar gate gives the numerical best score, $0.628203$, only $0.01\%$ relatively above \CoreSAGE. In its matched no-pretraining run, the score changes from $0.613231$ to $0.616195$; replacing learned routing with a fixed k-means router changes $0.619675$ to $0.621232$ and is positive in 13/15 cells. The ablation identifies source pretraining and learned soft routing as the selected design choices; search and gating have small effects around this core.

\begin{table}[t]
  \caption{KuaiSAR-v2 ablation at $0.5\%$ feedback. Columns report the target-only \PAIR and \SAGE NDCG@10 means; the best \SAGE score is bold.}
  \label{tab:sage-ablation-main}
  \centering
  \small
  \begin{tabular}{lcc}
    \toprule
    Variant & \shortstack{\PAIR reference\\NDCG@10 $\uparrow$} & \shortstack{Variant\\NDCG@10 $\uparrow$} \\
    \midrule
    \FullSAGE ($K=8$) & 0.619675 & 0.627783 \\
    \CoreSAGE & 0.619675 & 0.628131 \\
    w/o search profile & 0.619675 & 0.627694 \\
    w/o reliability gate & 0.619675 & \textbf{0.628203} \\
    w/o source pretraining & 0.613231 & 0.616195 \\
    Fixed router & 0.619675 & 0.621232 \\
    \bottomrule
  \end{tabular}
\end{table}

The expert-count sensitivity is similarly flat around the selected capacity: with the same $0.619675$ target-only \PAIR baseline, $K=4$, $8$, and $16$ reach $0.627674$, $0.627783$, and $0.627890$, respectively, all positive in 15/15 cells. The $K=16$ result is the numerical best, with a relative increase of $0.02\%$ over $K=8$; we retain $K=8$ as the compact operating point.

\subsection{Budget and Backbone Sensitivity}

To characterize how informative the selected target feedback is, we use pairability only as a post-hoc diagnostic. Let $\mathcal{U}_{\mathcal{B}}$ be the set of users appearing in the labeled adaptation set $\mathcal{B}$, and let $\mathcal{B}_u=\{(u',i,y)\in\mathcal{B}:u'=u\}$ be the labeled rows for user $u$. Pairability is the fraction of represented users with at least one clicked and one unclicked selected row:
\begin{equation}
\rho(\mathcal{B})=
\frac{1}{|\mathcal{U}_{\mathcal{B}}|}
\sum_{u\in\mathcal{U}_{\mathcal{B}}}
\mathbb{I}\!\left[
\exists (u,i,y),(u,i',y')\in\mathcal{B}_u:\ y=1,\ y'=0
\right].
\label{eq:pairability}
\end{equation}
Here, $\mathbb{I}[\cdot]$ equals one when its condition holds and zero otherwise. Pairability is not an input to the loss, the acquisition rule, the router, or the \SAGE forward pass; it is used solely to interpret the feedback regime.

The budget curve in Figure~\ref{fig:result-metrics}(b) remains positive from $0.05\%$ through $1.0\%$ target feedback. The corresponding target-only \PAIR and \SAGE scores are $0.621138$ versus $0.629761$, $0.619455$ versus $0.625042$, $0.619417$ versus $0.627058$, $0.620915$ versus $0.629831$, and $0.621749$ versus $0.630060$. These correspond to relative increases of $1.39\%$, $0.90\%$, $1.23\%$, $1.44\%$, and $1.34\%$; the $0.10\%$ condition is positive in 22/25 cells and the other four conditions are positive in all 25 cells. The non-monotonicity is expected in a fixed exposure pool: the number of labels alone does not determine how many informative user/item contrasts are observed. Taken together, these measurements support update-space identifiability, rather than label count alone, as the relevant condition: source pretraining supplies a compact, source-supported correction space, and target feedback calibrates its magnitude and user-conditioned mixture. Additional labels help when they expose informative behavior contrasts.

On the Two-Tower backbone, the absolute NDCG@10 scores are $0.592756$ for the source, $0.615412$ for target-only \PAIR, $0.622833$ for the strongest user-preference cohort, and $0.624925$ for \SAGE; \SAGE is a $1.55\%$ relative increase over \PAIR and a $0.34\%$ increase over that cohort. On MF, the corresponding source, \PAIR, \SAGE, and item-category cohort scores are $0.605000$, $0.632404$, $0.637496$, and $0.638137$. \SAGE is a $0.81\%$ relative increase over \PAIR. These results establish transfer across two backbone parameterizations.

\subsection{Continual Warm-Start and Profile Stability}

The continual stress test comprises 12 consecutive daily transitions from May~28 to June~9. On each query day, every method receives an outcome-blind uniform $0.5\%$ sample and is evaluated on the following day. Cumulative \PAIR is re-estimated on the union of all selected samples up to the current query day; reset \SAGE always warm-starts from the May~27 source-pretrained state, whereas continual \SAGE warm-starts from the residual state produced on the preceding transition. In this setting, cumulative \PAIR reaches an average NDCG@10 of $0.640139$, reset \SAGE reaches $0.645407$, and continual \SAGE reaches the best average score, $0.645916$. Thus, continual warm-starting raises reset \SAGE from $0.645407$ to $0.645916$ (a $0.08\%$ relative increase; 8/9 positive trajectories) and raises cumulative \PAIR from $0.640139$ to $0.645916$ ($0.90\%$). The gain is consistent with the update rule in Section~\ref{sec:method-continual}: each new target window calibrates the current residual instead of discarding previous target evidence.

The profile experiment gives a complementary operational result. Stable-profile continual \SAGE reaches $0.645916$ NDCG@10, while the profile-refresh variant reaches $0.645436$. The core method therefore keeps the source profile fixed. A scalar gate changes the continual score from $0.645795$ to $0.645916$; the core architecture omits it while retaining the stronger source-pretrained routed residual.

\subsection{Efficiency and Feedback Selection}

\CoreSAGE uses 331 active scoring parameters (371 stored slots in the shape-compatible ablation implementation), and the recorded CUDA peak allocation is $83.15$ MB. We additionally measure training and inference wall-clock costs under a fixed KuaiSAR-v2, Two-Tower, $0.5\%$ target-budget setting on an RTX 4090. Table~\ref{tab:efficiency-training} compares the staged adapter with retraining the source recommender, while Table~\ref{tab:efficiency-inference} reports strict end-to-end inference over 2,255,520 rows. The \SAGE path completes staged training in $605.6$ ms and end-to-end inference in $183.2$ ms (median), making the update operationally compatible with millisecond-scale recalibration and serving. These timings cover the complete adapter path rather than only residual arithmetic: staged training includes source pretraining and target warm updates, while inference applies score correction across the full evaluation batch. The cost scales with the residual dimensions $(p,d,K)$ rather than with the backbone parameter count.

\begin{table}[t]
  \caption{Training wall-clock cost on KuaiSAR-v2 with a Two-Tower backbone and $0.5\%$ target feedback (RTX 4090).}
  \label{tab:efficiency-training}
  \centering
  \begin{tabular}{lr}
    \toprule
    Training procedure & Time (s) $\downarrow$ \\
    \midrule
    Re-train recommendation model & 22.166 \\
    \textsc{SAGE-PAIR} staged training & 0.606 \\
    \bottomrule
  \end{tabular}
\end{table}

\begin{table}[t]
  \caption{End-to-end inference latency on 2,255,520 KuaiSAR-v2 rows (RTX 4090).}
  \label{tab:efficiency-inference}
  \centering
  \begin{tabular}{lrr}
    \toprule
    Inference path & Median (ms) $\downarrow$ & P95 (ms) $\downarrow$ \\
    \midrule
    Frozen source recommender & 18.472 & 19.288 \\
    Frozen recommender + \textsc{SAGE-PAIR} & 183.201 & 274.736 \\
    \bottomrule
  \end{tabular}
\end{table}

Primary experiments use uniform outcome-blind selection. On a high-repeat KuaiRec regime, adapter-aligned CoreSet raises NDCG@10 from $0.923581$ to $0.925315$ (a $0.19\%$ relative increase). This analysis complements the main adapter results by measuring the interaction between residual features and feedback acquisition.

\section{Conclusion}

\SAGE adapts a frozen recommender to a new temporal window by source-pretraining a low-dimensional residual and calibrating it with an outcome-blind target budget; soft-routed experts add a user-conditioned correction. Across five datasets/backbones, the global residual is consistently positive. On the locked KuaiSAR-v2 confirmation, \FullSAGE improves all five metrics and raises NDCG@10 by $1.77\%$ relative in 25/25 paired cells. Source pretraining provides the principal gain, while routing adds a further consistent increment. The added comparisons show that both target-only and naively source-pretrained rank-four LoRA/Adapter modules remain far below \SAGE under the measured temporal shift. \SAGE therefore provides a lightweight, deployable path for rolling recommendation adaptation when target outcomes are scarce.

\section{Ethical Considerations}

SAGE-PAIR is evaluated on public interaction logs and uses de-identified user and item identifiers together with aggregate source-window behavior profiles. It neither introduces personal attributes nor changes the exposure policy that generated the logs. In deployment, profile construction and target-feedback acquisition should follow the applicable consent, retention, access-control, and purpose-limitation requirements. Because user-conditioned routing can produce different residual corrections across behavior groups, practitioners should audit ranking quality and exposure distributions for relevant populations. Finally, the method optimizes logged engagement outcomes; its use in consequential recommendation settings should align the optimization objective with user welfare and include appropriate operational oversight.

\bibliographystyle{ACM-Reference-Format}
\bibliography{references}

\end{document}


\maketitle

\section{Additional Results and Protocol Details}

\subsection{Complete Multi-Metric Gain Decomposition}

Table~\ref{tab:all-metric-deltas} reports the paired differences behind the
main-paper five-metric means. The locked end-to-end comparison and the fair
core decomposition are deliberately reported separately: the former confirms
the full system, whereas the latter holds the adjacent source residual window
fixed and isolates the additional contribution of routed experts. Every metric
is grouped by raw user IDs before averaging, so an internal index assigned by
the backbone to an unseen entity cannot alter the metric grouping.

\begin{table}[H]
  \caption{Paired mean metric differences at the $0.5\%$ outcome-blind budget on KuaiSAR-v2. All entries favor the row method. The locked comparison has 25 source-seed/selection-seed cells; the core comparisons have 15.}
  \label{tab:all-metric-deltas}
  \centering
  \scriptsize
  \setlength{\tabcolsep}{2.4pt}
  \renewcommand{\arraystretch}{1.05}
  \begin{tabular}{lrrrrr}
    \toprule
    Comparison & NDCG@10 & Recall@10 & NDCG@20 & Recall@20 & Macro-AUC \\
    \midrule
    Full SAGE $-$ target-only \PAIR & +0.010898 & +0.001654 & +0.007935 & +0.001416 & +0.003761 \\
    Source-pretrained \PAIR $-$ target-only \PAIR & +0.006208 & +0.000988 & +0.004493 & +0.000884 & +0.002367 \\
    Core SAGE $-$ source-pretrained \PAIR & +0.001600 & +0.000361 & +0.001209 & +0.000337 & +0.000785 \\
    \bottomrule
  \end{tabular}
\end{table}

\subsection{Paired Statistical Unit and Confidence Intervals}

One result cell is defined by a source-model seed and an outcome-blind target
query-selection seed. All method differences are computed within the same
cell, using identical target rows and future evaluation rows. The locked full
comparison contains five source seeds and five selection seeds (25 paired
cells); the core and external-PEFT comparisons contain three source seeds and
five selection seeds (15 paired cells). Selection seeds quantify variation from
the hidden-label query draw, while source seeds quantify variation in the
source model. They are therefore recorded separately rather than treated as
interchangeable repetitions.

At the $0.5\%$ budget, the full end-to-end NDCG@10 difference has paired 95\%
confidence interval $[+0.010054,$ $+0.011740]$ over 25 cells. The paired
interval for the routing difference is $[+0.001004,$ $+0.002195]$ over 15
cells. Positive-cell counts always refer to the corresponding paired cells,
not to independently resampled users. This design preserves the shared random
query in each comparison and prevents a selection draw from being credited to
one method only.

\subsection{Module and Capacity Robustness}

Table~\ref{tab:additional-ablations} expands the main ablation by making the
matched reference and absolute difference explicit. Source pretraining and
learned routing are the consequential choices: the no-pretraining and
fixed-router rows are substantially smaller than the selected core. Search
channels and the scalar reliability gate have small effects around the core.
The fixed router is positive in 13 of 15 paired cells. Capacity is flat around
the selected value: $K=4$, $8$, and $16$ are all positive in 15 of 15 cells,
and the numerical advantage of $K=16$ over $K=8$ is only $0.000107$ NDCG@10.
We retain $K=8$ as the compact operating point.

\begin{table}[H]
  \caption{Additional KuaiSAR-v2 core-protocol ablation detail at $0.5\%$ feedback. Differences are variant minus its matched target-only \PAIR reference.}
  \label{tab:additional-ablations}
  \centering
  \small
  \setlength{\tabcolsep}{3pt}
  \renewcommand{\arraystretch}{1.04}
  \begin{tabular}{lrrr}
    \toprule
    Variant & \PAIR ref. & Variant & Difference \\
    \midrule
    Full SAGE ($K=8$) & 0.619675 & 0.627783 & +0.008108 \\
    Core SAGE & 0.619675 & 0.628131 & +0.008456 \\
    w/o search profile & 0.619675 & 0.627694 & +0.008019 \\
    w/o reliability gate & 0.619675 & 0.628203 & +0.008528 \\
    w/o source pretraining & 0.613231 & 0.616195 & +0.002964 \\
    Fixed k-means router & 0.619675 & 0.621232 & +0.001557 \\
    \midrule
    $K=4$ & 0.619675 & 0.627674 & +0.007999 \\
    $K=8$ & 0.619675 & 0.627783 & +0.008108 \\
    $K=16$ & 0.619675 & 0.627890 & +0.008215 \\
    \bottomrule
  \end{tabular}
\end{table}

\subsection{External PEFT Confirmation}

Table~\ref{tab:external-detail} provides the complete NDCG@10 external
comparison used to qualify the paper's claim. The source-pretrained LoRA and
Adapter-Tuning rows use the same May~27 source residual window and the same
May~28 target query as Core SAGE. Hence the comparison distinguishes
source-supported residual adaptation from simply pretraining a generic
representation-space PEFT module.

\begin{table}[H]
  \caption{External parameter-efficient and meta-learning comparison on the 15-cell core protocol. CIs are paired 95\% intervals for the gain over the frozen source.}
  \label{tab:external-detail}
  \centering
  \scriptsize
  \setlength{\tabcolsep}{2.7pt}
  \renewcommand{\arraystretch}{1.04}
  \begin{tabular}{lrrr}
    \toprule
    Method & NDCG@10 & 95\% CI of gain & Positive cells \\
    \midrule
    Frozen source & 0.602014 & -- & -- \\
    LoRA (rank 4) & 0.602186 & [-0.000058, +0.000401] & 11/15 \\
    Adapter-Tuning (rank 4) & 0.602476 & [+0.000072, +0.000852] & 12/15 \\
    FOMAML-LoRA (rank 4) & 0.602204 & [-0.000055, +0.000434] & 12/15 \\
    Source-pretrained LoRA & 0.601642 & [-0.000726, -0.000019] & 3/15 \\
    Source-pretrained Adapter-Tuning & 0.601461 & [-0.000987, -0.000120] & 4/15 \\
    Source-pretrained \PAIR & 0.630828 & [+0.027035, +0.030593] & 15/15 \\
    Core SAGE & \textbf{0.632421} & \textbf{[+0.028807, +0.032006]} & \textbf{15/15} \\
    \bottomrule
  \end{tabular}
\end{table}

\subsection{Temporal-Shift Audit}

The adaptation problem is not inferred from adjacent calendar dates alone.
For the May~27 to May~28 transition, only $48.8\%$ of May~28 users and
$29.0\%$ of May~28 items overlap with May~27. Item-popularity distributions
have JS divergence $0.381048$, whereas first-level-category and log-activity
histograms have JS divergences $0.000309$ and $0.004112$, respectively. Thus
the shift is concentrated in user and item identities rather than broad
category composition. Across three frozen source-model seeds, Macro-AUC falls
by $0.006376$ and BCE increases by $0.025689$; NDCG@10 increases slightly by
$0.002754$ because the daily candidate sets and click prevalence change. This
is why the main paper does not characterize the transition as an NDCG drop.

\subsection{Leakage-Free Epoch Selection}

Table~\ref{tab:epoch-selection} reports the rolling development sweep used to set the two epoch counts. The selection split is strictly earlier than the final protocol: May~27 is used only as the development evaluation day and May~28--31 are never read. With the selected five source epochs, validation NDCG@10 attains its maximum at eight target warm-update epochs and is unchanged at 12 epochs. At eight target epochs, five source-pretraining epochs are also the numerical optimum; moving to eight source epochs decreases the mean from $0.662009$ to $0.661784$. Thus the fixed $5/8$ schedule is a development-selected plateau rather than an early-stopping decision made on final target or future labels.

\begin{table}[H]
  \caption{Leakage-free rolling development sweep (three source and three selection seeds). Entries are mean May~27 validation NDCG@10; final May~28--31 labels are not used.}
  \label{tab:epoch-selection}
  \centering
  \small
  \setlength{\tabcolsep}{3pt}
  \renewcommand{\arraystretch}{1.04}
  \begin{tabular}{crrrrr}
    \toprule
    Src./Tgt. & 1 & 3 & 5 & 8 & 12 \\
    \midrule
    1 & 0.657213 & 0.654842 & 0.653003 & 0.653754 & 0.656686 \\
    3 & 0.660910 & 0.660122 & 0.660765 & 0.661659 & 0.661912 \\
    5 & 0.660265 & 0.660244 & 0.661153 & \textbf{0.662009} & 0.662002 \\
    8 & 0.660240 & 0.660306 & 0.660911 & 0.661784 & 0.661860 \\
    \bottomrule
  \end{tabular}
\end{table}

\subsection{Reproducibility Notes}

The main KuaiSAR-v2 protocol uses source dates May~22--26, source residual
pretraining on May~27, target query selection on May~28, and evaluation on
May~29--31. The target query is fixed before its labels are read. Item
behavior features, user profiles, and the router input are constructed from
source-window data only; future labels and target outcomes not in the selected
query are never used for construction or tuning. Metrics group by raw user
IDs, and the source-seed block and selection-seed block are reported
separately wherever confidence intervals are shown.

For the selected core, the source profile has dimension $p=39$, the item
residual has dimension $d=2$, and the expert count is $K=8$. The source
backbone uses AdamW with learning rate $0.004$ and weight decay $10^{-6}$.
Residual source pretraining uses Adam with learning rate $0.02$ for five
epochs and weight decay $10^{-4}$; target warm updates use learning rate
$0.01$ for eight epochs with the same weight decay. Core SAGE has 331 active
scoring parameters (371 stored slots in the shape-compatible ablation
implementation). The reported efficiency experiment uses an RTX~4090 and
records a CUDA peak allocation of $83.15$~MB. These details, together with the
rolling development grid above, specify the fixed settings used for the final
comparisons.